\documentclass[conference]{IEEEtran}

\usepackage{paralist, tabularx}
\usepackage{latexsym}
\usepackage{graphicx}
\usepackage{listings}
\usepackage{algorithm}
\usepackage{algpseudocode}
\usepackage{booktabs}
\usepackage{multirow}
\usepackage{siunitx}
\usepackage{comment}
\usepackage{amsmath}
\usepackage{amssymb}
\usepackage{color}
\usepackage{xspace}
\usepackage{subcaption}
\usepackage{tikz}
\usepackage{pgfplots}
\usepackage{breakurl}
\usepackage{url}

\usepackage{flushend}

\newtheorem{example}{Example}

\newcommand{\toolname}{{\sc Astroid}\xspace}
\newcommand{\apposcopy}{{\sc Apposcopy}\xspace}
\newcommand{\drebin}{{\sc Drebin}\xspace}
\newcommand{\massvet}{{\sc MassVet}\xspace}

\newcommand{\wbo}{{\sc Open-WBO}\xspace}

\newcommand*{\comments}{}
\ifdefined\comments
\newcommand{\yu}[1]{\textcolor{blue}{\textbf{YU:} #1}}
\newcommand{\osbert}[1]{\textcolor{green}{\textbf{OSBERT:} #1}}
\newcommand{\isil}[1]{\textcolor{red}{\textbf{I\c{S}IL:} #1}}
\newcommand{\todo}[1]{\textcolor{red}{#1}}
\else
\newcommand{\yu}[1]{}
\newcommand{\osbert}[1]{}
\newcommand{\isil}[1]{}
\newcommand{\todo}[1]{}
\fi

\begin{document}
\title{Automated Synthesis of Semantic Malware Signatures using Maximum Satisfiability}

\author{
\IEEEauthorblockN{Yu Feng}
\IEEEauthorblockA{
University of Texas at Austin\\
yufeng@cs.utexas.edu}
\and
\IEEEauthorblockN{Osbert Bastani}
\IEEEauthorblockA{
Stanford University\\
obastani@cs.stanford.edu
}\\[0.2cm]
\IEEEauthorblockN{\hspace*{-6cm}Isil Dillig}
\IEEEauthorblockA{
\hspace*{-6cm}University of Texas at Austin\\
\hspace*{-6cm}isil@cs.utexas.edu}
\and
\IEEEauthorblockN{Ruben Martins}
\IEEEauthorblockA{
University of Texas at Austin\\
rmartins@cs.utexas.edu
}\\[0.2cm]
\IEEEauthorblockN{\hspace*{-6cm}Saswat Anand}
\IEEEauthorblockA{
\hspace*{-6cm}Google, Inc.\\
\hspace*{-6cm}saswatanand@google.com}
}


%Telephone: (800) 555--1212, Fax: (888) 555--1212}


\IEEEoverridecommandlockouts
\makeatletter\def\@IEEEpubidpullup{9\baselineskip}\makeatother
\IEEEpubid{\parbox{\columnwidth}{Permission to freely reproduce all or part
    of this paper for noncommercial purposes is granted provided that
    copies bear this notice and the full citation on the first
    page. Reproduction for commercial purposes is strictly prohibited
    without the prior written consent of the Internet Society, the
    first-named author (for reproduction of an entire paper only), and
    the author's employer if the paper was prepared within the scope
    of employment.  \\
    NDSS '17, 26 February - 1 March 2017, San Diego, CA, USA\\
    Copyright 2017 Internet Society, ISBN 1-1891562-46-0\\
    http://dx.doi.org/10.14722/ndss.2017.23379
}
\hspace{\columnsep}\makebox[\columnwidth]{}}

\maketitle
\begin{abstract}
This paper proposes a technique for automatically learning semantic malware signatures for Android from very few samples of a malware family. The key idea underlying our technique is to look for a \emph{maximally suspicious common subgraph (MSCS)} that is shared between all known instances of a malware family. An MSCS describes the shared functionality between multiple Android applications in terms of  inter-component call relations and their semantic metadata (e.g., data-flow properties). Our approach identifies such maximally suspicious common subgraphs by reducing the problem to \emph{maximum satisfiability}. Once a semantic signature is learned,  our approach uses a combination of static analysis and a new \emph{approximate signature matching} algorithm to determine whether an Android application matches the semantic signature characterizing a given malware family.

We have implemented our approach in a tool called \toolname
and show that it has a number of advantages over state-of-the-
art malware detection techniques. First, we compare the semantic
malware signatures automatically synthesized by \toolname with
manually-written signatures used in previous work and show
that the signatures learned by \toolname perform better in terms
of accuracy as well as precision. Second, we compare \toolname against two 
state-of-the-art malware detection tools and demonstrate its advantages in
terms of interpretability and accuracy. Finally, we demonstrate that
\toolname's approximate signature matching algorithm is resistant to
behavioral obfuscation and that it can be used to detect zero-day
malware. In particular, we were able to find 22 instances of zero-day
malware in Google Play that are not reported
as malware by existing tools.
\end{abstract}

\section{Introduction}\label{sec:intro}

Due to the enormous popularity of Android as a mobile platform, the number of applications (``apps") available for Android has skyrocketed, with 1.6 million apps being currently available for download. Unfortunately, the soaring number of Android users has also led to a rapid increase in the number of Android malware, with 4,900 malware samples being introduced \emph{every day}~\cite{gdata-report}. 
Correspondingly, this upsurge in Android malware has also led to a flurry of research for automatically detecting malicious applications~\cite{apposcopy,drebin,DroidSift,Bose08,holmes,DroidRanger}. 

Generally speaking, approaches for automated malware detection can be classified as either \emph{signature-based} or \emph{learning-based}.
Signature-based techniques look for specific patterns in the application to determine whether the app is malicious, and, if so, which malware family the app belongs to~\cite{Kirin,apposcopy,Hancock,SB05}. These  patterns can either be syntactic (e.g., sequence of instructions) or semantic (e.g., control- or data-flow properties). Signature-based approaches allow security analysts to quickly identify the malicious component of an application; hence, they are widely-used by several commercial anti-virus (AV) companies. However, one key shortcoming of these techniques is that they require a trained security analyst to \emph{manually} write suitable signatures that can be used to detect each malware family. Unfortunately, this manual effort is typically time-consuming and error-prone.

\emph{Learning-based} techniques~\cite{drebin,holmes,smit,aisec,DroidSift,peng12,DroidMiner,Bose08} aim to address this limitation by automatically learning a malware classifier from data. These techniques  extract various features from the application  and use standard machine learning algorithms to learn a classifier that labels apps as either benign or malicious. 
Compared to signature-based techniques, current learning-based approaches suffer from a number of shortcomings:
\begin{itemize}
\item They produce results that are difficult to interpret (for example, they typically cannot be used to determine the malware family), which makes it difficult for a security analyst to discharge false positives.
\item They typically require a large number of  samples from each malware family, which is problematic for families that have recently emerged or that are rare (58\% of malware families in~\cite{drebin} have fewer than 5 samples).
\end{itemize}

This paper aims to overcome these disadvantages of existing malware detectors by proposing a new technique to \emph{automatically infer malware signatures}. By identifying malware based on inferred signatures, our approach retains all the advantages of signature-based approaches: it can pinpoint the location of the malicious components as well as the corresponding malware family, and requires very few samples ($<$5 in our evaluation). Furthermore, our approach can learn \emph{semantic} signatures that are resilient to low-level obfuscation mechanisms  and produces very few false alarms.
Finally, because our signature matching algorithm uses \emph{approximate} (rather than \emph{exact}) matching, our algorithm is also resilient to high-level obfuscation mechanisms that modify the program's control flow and data-flow properties.

The first key insight underlying our approach is to automatically identify a semantic pattern that (a) occurs in all instances of a given malware family, and (b) is \emph{maximally suspicious} (i.e., maximizes the number of ``suspicious" features that are  typically \emph{not} found in benign apps). Here, criterion (a) serves to minimize the number of false negatives: If the pattern does not occur in \emph{all} instances of the malware family, then our signatures would not match all malicious apps, thereby resulting in false negatives. In contrast, criterion (b) serves to minimize false positives: By requiring that the identified semantic pattern is \emph{maximally suspicious}, we ensure that our signatures are unlikely to flag benign apps as malicious.

The second key insight underlying our technique is to automatically learn these semantic patterns by finding a \emph{maximally suspicious common subgraph (MSCS)} of the malware instances. An MSCS describes the shared functionality between multiple Android applications in terms of  inter-component call relations and their semantic metadata (e.g., data-flow properties, intent filters, API calls etc.). Our approach automatically finds an MSCS  by reducing the problem to \emph{maximum satisfiability (MaxSAT)}. Intuitively, the MaxSAT problem aims to maximize the amount of suspicious metadata while ensuring that this functionality is shared between all instances of the malware family. The solution to the MaxSAT problem can then be directly translated into a semantic signature that characterizes a specific malware family.

Our third and final insight is to utilize the proposed signature inference algorithm for approximate matching. Specifically, given 
a database of malware signatures  and a new application $\mathcal A$, we must decide whether $\mathcal A$ matches any of these signatures. Rather than performing \emph{exact} signature matching as done in previous work~\cite{apposcopy}, we employ a novel \emph{approximate signature matching} algorithm. The key idea is to generate a new signature $\mathcal S$ \emph{assuming} that  $\mathcal A$  is an instance of  family $\mathcal F$. We then decide if  $\mathcal A$ is actually an instance of $\mathcal F$ based on the similarity score between $\mathcal S$ and $\mathcal F$'s existing signature. The main advantage of this approach is that it makes our technique even more resilient to obfuscations, including those that change program behavior.

We have implemented the proposed technique in a tool called \toolname\footnote{\toolname \ stands for \emph{Automatic SignaTure findeR for andrOID}} and evaluate it in a number of ways. Our first experiment shows that the signatures automatically inferred by \toolname \ are competitive with (in fact, better than) manually-written signatures used for evaluating previous work~\cite{apposcopy}. We also compare \toolname \ with two state-of-the-art Android malware detectors, namely \drebin~\cite{drebin} and \massvet~\cite{MassVet}, and show that \toolname compares favorably with these tools in terms of accuracy, false positives, and interpretability. Third, we  demonstrate that \toolname's approximate signature matching algorithm is both resilient to various behavioral obfuscations and is also useful for detecting zero-day malware. Specifically, in a corpus of apps collected from Google Play, \toolname identified 103 apps from more than 10 malware families for which we did not previously have signatures. Furthermore, among these 103 apps, 22 of them were previously unknown  and are not reported as malware by existing tools.

This paper makes the following key contributions:

\begin{itemize}
\item We propose a novel technique for inferring malware signatures from few samples of a  malware family.
\item We formulate signature inference as the problem of finding a \emph{maximally suspicious common subgraph (MSCS)}  of a set of Android applications and show how to reduce MSCS detection to \emph{maximum satisfiability}.
\item We propose a novel \emph{approximate signature matching} algorithm that leverages automated signature synthesis.
\item We implement our approach in a tool called \toolname \ and
evaluate it against manually written malware signatures, other state-of-the-art malware detectors, and behavioral obfuscation strategies. 
\end{itemize}

\begin{figure*}[!t]
%\vspace{-0.2in}
\centering
\includegraphics[scale=0.21]{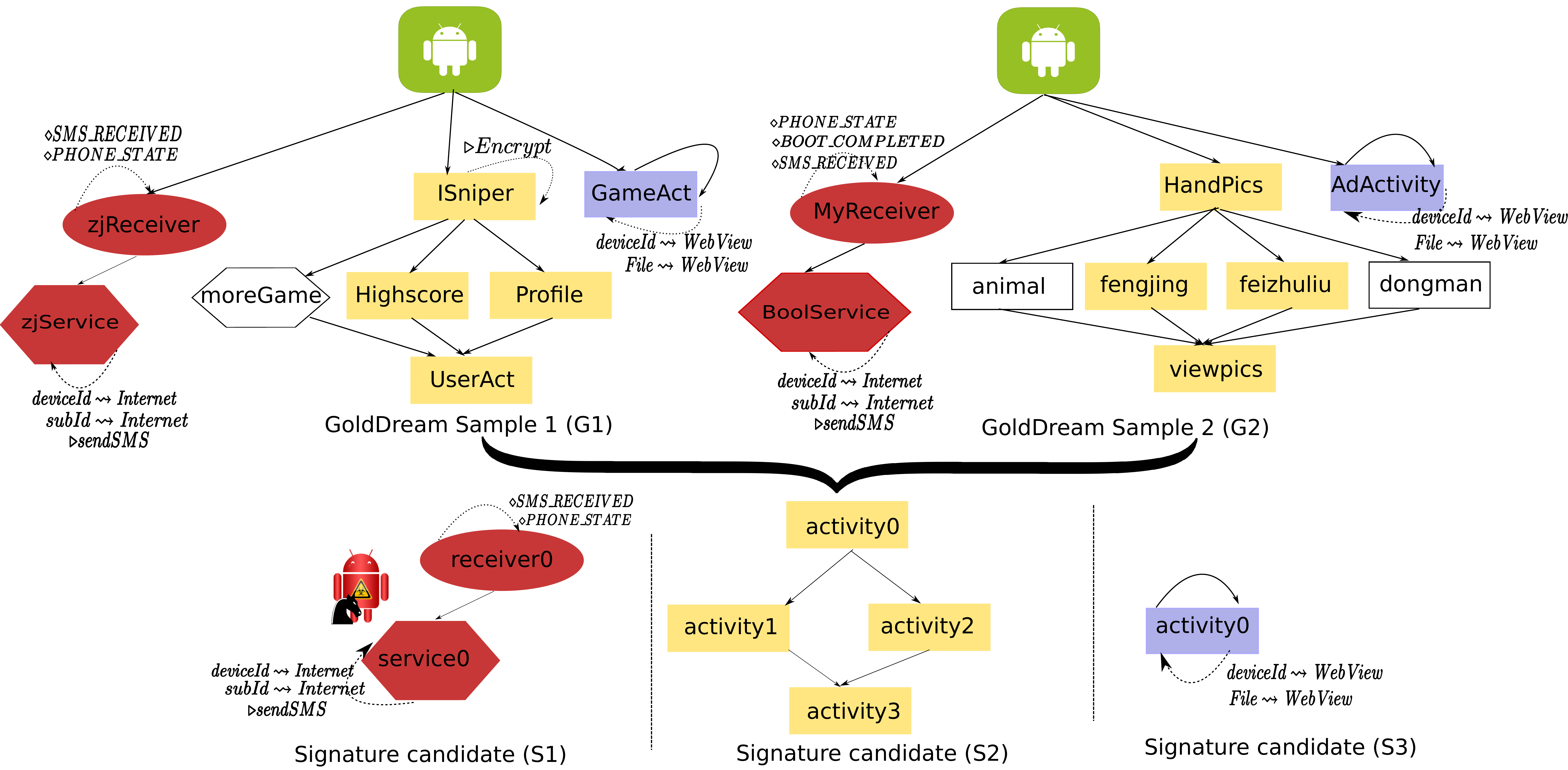}
\caption{Motivating example to illustrate our approach}\label{fig:golddream}
%\vspace*{-4mm}
\end{figure*}

\section{Background}\label{sec:background}

In this section, we provide some background that is necessary for understanding the rest of this paper. 

\subsection{Android Basics}

An Android application consists of four kinds of components, namely \emph{Activity},
\emph{Service}, \emph{BroadcastReceiver}, and \emph{ContentProvider}. Every screen of an
 app corresponds to an Activity. Services run in the
background without a user interface, ContentProviders store data, and BroadcastReceivers
react asynchronously to messages from other applications. 

In Android, different components communicate with each other through \emph{Intents}, which are effectively messages that describe  operations to be performed. Components can receive intents from other components or from
 the Android system; thus intents form the basis of 
 \emph{inter-component communication (ICC)} in Android.  Intent objects can have several attributes, 
 three of which we discuss below:
\begin{compactitem}
\item The optional {\bf \emph{target}} attribute explicitly specifies 
the specific component invoked by this Intent.
\item The {\bf \emph{action}} attribute specifies the type of
 action that the receiver of the Intent should perform.
\item The {\bf \emph{data type}} attribute
 specifies the type of data that the receiver of the intent is supposed to operate on.
\end{compactitem}
An Intent whose target attribute is specified is called an
\emph{explicit Intent}. In contrast, an \emph{implicit Intent} does not have its target
attribute  specified, so the Android system decides the targets of  an implicit intent $I$ at run time by
comparing $I$'s action and data attributes  against the \emph{intent filters} of other components.
The intent filter of a component $C$ specifies the action that $C$ is able to perform and the data it can operate on.
Such intent filters are declared in the app's \emph{manifest}, which is an XML file containing information
about the app.

\subsection{Inter-component Call Graphs}

Our technique uses  the \emph{inter-component call graph (ICCG)} representation introduced in
previous work~\cite{apposcopy}. The ICCG for an Android app  summarizes  inter-component communication
within the app as well as any relevant metadata (e.g., dataflow or API calls). 
More formally, an ICCG for an application $A$ is a graph $(V, X, Y)$ where:

\begin{compactitem}
\item $V$ is a set of vertices, where each $v \in V$  is a component of $A$.
The \emph{type} of a vertex $v$, written $T(v)$, is the type of the component that $v$ represents. Hence, we have:
\[
T(v) \in \mathcal{T} = \{ activity, service, receiver, provider \}
\]
\item $X$ is a set of edges representing inter-component call relations. Specifically, 
$(v, v') \in X$ indicates that component $v$ may invoke component $v'$ either through an explicit or implicit intent.
\item $Y$ is a set of labeled edges representing metadata. In particular, $(v, v', d) \in Y$ indicates that 
 components $v$ and $v'$ are related by metadata $d$.
\end{compactitem}

\vspace{0.05in}
Metadata edges $Y$ in the ICCG  indicate 
potentially suspicious behaviors of the app. We explain three kinds 
of metadata that will be used in this paper:

\begin{itemize}
\item \emph{\bf \emph{Data flow information}}, denoted as $\emph{src} \leadsto \emph{sink}$. 
In particular, $(v, v', s \leadsto s') \in Y$ indicates that the source \emph{s} 
originating from component $v$ flows to the sink $s'$ in component $v'$. Sources represent confidential
information (e.g.,  IMEI number), and sinks represent externally  visible channels (e.g., \emph{Internet}, \emph{SMS}).
\item {\bf \emph{Suspicious API calls}}, written  $\triangleright \emph{API}$. Specifically, an edge $(v, v, \triangleright \emph{f}) \in Y$ indicates 
component $v$ calls Android API method $f$ (e.g., {\tt exec}).
\item {\bf \emph{Suspicious actions}}, denoted as $\diamond \emph{Act}$, represent actions that can be performed by a component. In particular, $(v, v, \diamond \emph{Act}) \in Y$ indicates that component $Y$ can perform action $\emph{Act}$.
\item {\bf \emph{Intent filters}}, denoted as $\star o$, represent the data type that a given component can operate on. In particular, $(v, v, \star o) \in Y$ indicates that component $v$ can operate on data of type $o$ (passed using intents).
\end{itemize}
 
For a given set of apps, the universe of labels from which  metadata can be drawn is fixed. In the rest of the paper, we  use  $\mathcal{D}$ to indicate the domain of edge labels $d$ in $Y$.

\begin{example}
Consider the ICCG labeled as \emph{Gold Dream Sample 1} in Figure~\ref{fig:golddream}. Here,  solid edges represent $X$, and dashed edges  with metadata represent $Y$. Nodes drawn as rectangles are Activities, while ellipses indicate BroadcastReceivers and hexagons denote Services. Since there is a solid edge from {\tt zjReceiver} to {\tt zjService}, broadcast receiver {\tt zjReceiver} may call {\tt zjService}. Furthermore, since there is a dashed edge from {\tt zjService} to itself labeled as ${\rm deviceId} \leadsto \emph{Internet}$,  source \emph{deviceId} flows to the Internet within component {\tt zjService}.
\end{example}

\section{Overview}\label{sec:sol}

Suppose that Alice, a security auditor at an anti-virus company, recently learned about a new malware family called GoldDream~\cite{JiangDream}. Alice would like to update their anti-virus tool to detect GoldDream instances. Alice wants to use a learning-based tool to save work, but then she must manually examine each app flagged by the tool due to high false positive rates (which is time-consuming because the explanations produced by these tools typically do not pinpoint malicious functionalities). Furthermore, the malware family is new, so Alice only has two samples, and current learning-based approaches require many training samples to achieve high accuracy. Defeated, Alice manually writes signatures to detect GoldDream instances by reverse engineering common malicious behaviors from the byte code of the two GoldDream samples, and working to ensure that no benign apps are flagged by her signatures.

\toolname can greatly benefit Alice by automatically inferring a signature characterizing the GoldDream family from as few as two samples. In the top half of Figure~\ref{fig:golddream}, we show the ICCGs of Alice's two GoldDream samples. Observe that these two samples have different component names and perform very different functionalities---the ICCG on the left belongs to a game, while the one on the right belongs to an app for browsing pictures.

To detect the malice shared by these two samples, \toolname searches for {connected} subgraphs of the two ICCGs that are isomorphic to each other. In this case, there are multiple such subgraphs; three are shown in the bottom half of Figure~\ref{fig:golddream}. For instance, the red graph labeled \emph{Signature candidate (S1)} is isomorphic to the red subgraphs of $G1$ and $G2$. Similarly, the yellow graph labeled \emph{Signature candidate (S2)} is isomorphic to the yellow subgraphs of $G1$ and $G2$. The key insight in deciding between these candidates is to find the \emph{maximally suspicious common subgraph (MSCS)}. Intuitively, an MSCS is the signature candidate that maximizes the number of metadata edges, where each edge is weighted by its suspiciousness.

Continuing our example, \emph{S2} does not exhibit any ``suspicious'' behaviors such as suspicious information flows from confidential sources to public sinks. On the other hand, \emph{S1} contains multiple suspicious behaviors, such as calling the \emph{sendSMS} API and leaking confidential data. Based on these suspicious features, \toolname decides that among the three candidates \emph{S1-S3}, the candidate \emph{S1} most likely encodes the malicious behavior characterizing malware in the GoldDream family.

Internally, \toolname uses a MaxSAT solver to find MSCSs. Since each suspicious behavior is encoded as a \emph{soft constraint}, an optimal satisfying assignment to the MaxSAT problem corresponds to a maximally suspicious common subgraph of the malware samples. Once \toolname infers an MSCS of the malware samples, it automatically converts this MSCS into a signature. Hence, \toolname allows a security auditor like Alice to \emph{automatically} detect future instances of the GoldDream family without having to manually write malware signatures.

\section{Semantic Android Malware Signatures}
\label{sec:prob}

We now formally define our \emph{malware signatures} and state what it means for an app to \emph{match} a signature.
Intuitively, a signature for a family $\mathcal{F}$ is an ICCG $(V_0, X_0,Y_0)$
that captures  semantic properties common to all malware in $\mathcal{F}$.
Ideally, $G_0 = (V_0, X_0,Y_0)$ would satisfy the following:
\begin{itemize}
\item $G_0$ occurs as a subgraph (defined below) of the ICCG $G_S = (V_S, X_S,Y_S)$ of every malware sample $S\in\mathcal{F}$.
\item $G_0$ does \emph{not} occur as subgraph of the ICCG $G_S = (V_S, X_S,Y_S)$ of any sample $S\not\in\mathcal{F}$.
\end{itemize}
By ``occurs as a subgraph'', we mean  there exists an embedding
$F_S:V_0\to V_S$
such that the following properties hold:
\begin{itemize}
\item {\bf One-to-one.} For every $v,v'\in V_0$ where $v\not=v'$, $F_S$ cannot map both $v$ and $v'$ to the same vertex, i.e.,
$
F_S(v)\not=F_S(v').
$
\item {\bf Type preserving.} For every $v\in V_0$, $F_S$ must map $v$ to a vertex of the same type, i.e., $
T(v)=T(F_S(v))$
\item {\bf Edge preserving.} For every $v,v'\in V_0$, $F_S$ must map an edge $(v,v')\in X_0$ to an edge in $X_S$:
\begin{align*}
(v,v')\in X_0\Rightarrow (F_S(v),F_S(v'))\in X_S.
\end{align*}
\item {\bf Metadata preserving.} For every $v,v'\in V_0$, $F_S$ must map metadata $(v,v',d)\in Y_0$ to metadata in $Y_S$:
\begin{align*}
(v,v',d)\in Y_0\Rightarrow (F_S(v),F_S(v'),d)\in Y_S.
\end{align*}
\end{itemize}

\begin{example}
\label{ex:prob}
Consider the candidate signature $S1$ with ICCG $(V_0,X_0,Y_0)$ and the GoldDream sample $G1$ with ICCG $(V_1,X_1,Y_1)$ from Figure~\ref{fig:golddream}. Now, let us consider the following candidate embeddings $F_1^{(a)}$ and $F_1^{(b)}$ from $V_0$ to $V_1$:
\begin{align*}
&\begin{cases}
F_1^{(a)}(\texttt{receiver0})=\texttt{zjReceiver} \\
F_1^{(a)}(\texttt{service0})=\texttt{zjService}
\end{cases} \\
&\begin{cases}
F_1^{(b)}(\texttt{receiver0})=\texttt{zjReceiver} \\
F_1^{(b)}(\texttt{service0})=\texttt{GameAct}.
\end{cases}
\end{align*}
Here, candidate $F_1^{(b)}$ is \emph{not} a valid embedding because the types of \texttt{service0} and \texttt{GameAct} are not compatible, since \texttt{service0} is a service whereas \texttt{GameAct} is an activity. Thus, $F_1^{(b)}$ does not preserve types.
Furthermore, $F_1^{(b)}$ is also invalid for another reason: There is an edge $\texttt{receiver0}\to\texttt{service0}\in X_0$ in $S1$,
but there is no corresponding edge $\texttt{zjReceiver}\to\texttt{GameAct}$ in  $G1$. Therefore, $F_1^{(b)}$ also does not preserve edges.
On the other hand, $F_1^{(a)}$ satisfies all the properties above and is a valid embedding.
\end{example}

Given signature $G_0 = (V_0, X_0, Y_0)$ and  app $S$ with ICCG $G_S = (V_S, X_S, Y_S)$,  we say that $G_0$ \emph{exactly matches} (or simply \emph{matches}) $S$ if $G_0$ occurs as a subgraph of $G_S$. In other words, given a signature $(V_0,X_0,Y_0)$ and a sample $S$ with ICCG $(V_S,X_S,Y_S)$, we can check whether $(V_0,X_0,Y_0)$
matches $S$. If so, we have determined that $S\in\mathcal{F}$; otherwise, $S\not\in\mathcal{F}$.

In general, our signature may not exactly capture the semantic properties of the malware family $\mathcal{F}$.
As a consequence, there may be samples $S\in\mathcal{F}$ such that $(V_0,X_0,Y_0)$ does not match $S$ (called \emph{false negatives}),
or samples $S\not\in\mathcal{F}$ such that $(V_0,X_0,Y_0)$ matches $S$ (called \emph{false positives}).
In practice, a signature should minimize  both the false positive rate and the false negative rate,
even though there is a tradeoff between optimizing these two values in practice.
When detecting malware, we use \emph{approximate matching}, which enables our tool to detect partial matches; tuning the cutoff for what constitutes an approximate match allows us to balance the tradeoff as desired. See Section~\ref{sec:approximate} for details.

\section{Signature Inference Problem}\label{sec:prob2}

Our goal is to \emph{infer} a signature from few samples of a malware family.
Suppose we are given $n$ malware samples from a single  family $\mathcal{F}$.
Na\"{i}vely, we can search for any signature that matches each of the $n$ malware samples;
then, the resulting signature would intuitively have a low false negative rate.
However, even the empty signature fits this criterion (since it can be embedded in any sample),
but the empty signature has a high false positive rate.

Instead, we want to maximize the amount of semantic information contained in the
signature.
More precisely, we seek to find a signature $(V_0,X_0,Y_0)$ that is:
\begin{itemize}
\item {\bf A common subgraph:} $(V_0, X_0, Y_0)$  should match each given sample $i\in\{1,...,n\}$. Intuitively, the common subgraph requirement seeks to minimize false negatives. If the inferred signature $S$ was not a common subgraph of all the samples, then $S$ would not match some samples of $\mathcal{F}$, meaning that we have false negatives. \\
\item {\bf Maximally suspicious:} $(V_0,X_0,Y_0)$ has maximal suspiciousness $|X_0|+\sum_{y\in Y_0}w_y$, where  weight $w_y$ indicates indicates the relative importance of  metadata edge $y$. Intuitively, \emph{maximal suspiciousness} seeks to minimize false positives: The less frequently a metadata edge $y$ appears in benign apps (i.e., the more ``suspicious" $y$ is), the higher the weight associated with it. Hence, maximizing the suspiciousness score  $ \sum_{y\in Y_0}w_y$ makes it less likely that a benign app will match the inferred signature. We describe how the weights $w_y$ are chosen in Section~\ref{subsec:function}.
\end{itemize}

In summary, given a set of samples $S_\mathcal{F}$ of malware family $\mathcal{F}$, we   refer to the problem of computing a signature $(V_0, X_0,Y_0)$ that is both maximally suspicious and a common subgraph of all  $G \in S_\mathcal{F}$ as  \emph{signature synthesis}.

\section{Signature Synthesis as MaxSAT}
\label{sec:sat}

As mentioned earlier, our approach reduces the signature synthesis problem to \emph{MaxSAT}~\cite{maxsat}.
Given a boolean formula in conjunctive normal form (CNF), the MaxSAT problem is to
find satisfying assignments to the variables in the formula that
maximizes the number of clauses that evaluate to true. For example, given the unsatisfiable formula
\begin{align*}
(x_0\vee x_1)\wedge(\neg x_0\vee x_1)\wedge(x_0\vee\neg x_1)\wedge(\neg x_0\vee\neg x_1),
\end{align*}
the assignment $\{x_0\mapsto0,x_1\mapsto0\}$ achieves the maximum of
three satisfied clauses.

In addition, we can specify that certain clauses \emph{must} be satisfied;
these clauses are referred to as \emph{hard constraints} and the remaining clauses
are referred to as \emph{soft constraints}.
We encode the common subgraph requirement of the signature synthesis problem using hard constraints. In contrast,  since the maximally suspicious requirement corresponds to optimizing an objective function, we encode this requirement using soft constraints  that should be maximally satisfied.

\subsection{Variables in Encoding}

Before we describe how to reduce the signature synthesis problem to MaxSAT, we introduce some terminology and 
describe the propositional variables used by our encoding.

First, we denote the ICCG of a given sample $i\in\{1,...,n\}$ of  malware family $\mathcal{F}$ as $(V_i,X_i,Y_i)$.
Recall that the common subgraph criterion requires  that the signature $(V_0,X_0,Y_0)$ matches each $i$ (i.e., we can find isomorphic embeddings $F_i:V_0\to V_i$).
The first difficulty in encoding the common subgraph requirement 
is that the number of vertices $|V_0|$ in the signature is unknown. However, for each
type $r\in\mathcal{T}$, we know that $V_0$ cannot contain more vertices of type $r$
than any sample (since it must be embedded in each sample).
Hence, we can give an upper bound on the number of vertices of type $r$ in $V_0$:
\begin{align*}
|\{v\in V_0\mid T(v)=r\}|\le m_r.
\end{align*}
Here, $m_r$ denotes the minimum number of vertices of type $r$ in any sample. This observation immediately gives us an upper bound on the total number of vertices:
\begin{align*}
|V_0|\le m=\sum_{r\in\mathcal{T}}m_r.
\end{align*}

Our approach is to fix a large set of vertices $V$, and then think of $V_0$ as an unknown subset of $V$ in our encoding.
Since there are at most $m$ vertices in the signature, it suffices to take $V=\{v_1,...,v_m\}$.
Furthermore, for each type $r$, we assign type $r$ to  exactly $m_r$ of the vertices in $V$.
Now, we can  think of each embedding $F_i$ as a \emph{partial} function $F_i:V\to V_i$, where we require that
the domain of $F_i$
is the same for each sample $i$.
Then, $V_0$ is exactly the \emph{common domain} of the $F_i$'s, and we have $X_0\subseteq V_0\times V_0$ as well
as $Y_0 \subseteq V_0 \times V_0 \times \mathcal{D}$.

\begin{example} 
Consider the two GoldDream samples shown in Figure~\ref{fig:golddream}. The first sample
has only 6 activities, 1 service, and 1 receiver, so $m_{\text{activity}}=6$, $m_{\text{service}}=m_{\text{receiver}}=1$, and
\begin{align*}
m=m_{\text{activity}}+m_{\text{service}}+m_{\text{receiver}}=8.
\end{align*}
Therefore, we take $V=\{v_1,...,v_8\}$, with 6 activities $v_1,...,v_6$, 1 service $v_7$, and 1 receiver $v_8$.
Note that the vertices $V_0$ used in the candidate signature must be a subset of $V$, and $X_0$ must be a subset of $V_0 \times V_0$.
\end{example}

In our description of the variables in our encoding, we use indices $i,j\in\{1,...,n\}$ to enumerate over the samples
and $k,h\in\{0,1,...,n\}$ to enumerate over both the samples and the signature.
Also, we use the indicator function:
\begin{align*}
\mathbb{I}[\mathcal{C}]=\begin{cases}1\text{ if }\mathcal{C}\text{ holds}\\0\text{ otherwise}\end{cases}.
\end{align*}

For readability, we use the notation $x(a,b,...)$ to denote a boolean variable $x_{a,b,...}$ indexed over $a\in A$, $b\in B$, and so forth.
Our constraint system is defined using the following constants and free variables:
\begin{itemize}
\item {\bf Domain indicators.} For every $v\in V$, a boolean variable  $d(v)$ indicates whether $v$ is in the domain of the embeddings $F_i$.
These are all free variables.
\item {\bf Embedding indicators.} For each $v\in V$,
\begin{align*}
f_i(v,u)=\mathbb{I}[F_i(v)=u].
\end{align*}
In other words, $f_i(v,u)$ indicates whether the embedding $F_i$ maps $v$ to $u$. These are all free variables.
\item {\bf Type indicators.} For each $v\in V_k$ (or $V$ if $k=0$) and $r\in\mathcal{T}$,   $t_k(v,r)$ indicates whether the type of $v$ is $r$:
\begin{align*}
t_k(v,r)=\mathbb{I}[T(v)=r].
\end{align*}
Since the types of all components are known, each $t_k(v, r)$ is a boolean constant.
\item {\bf Control-flow indicators.} For each $v,u\in V_k$ (or $V$ if $k=0$), $x_k(v,u)$ indicates whether there is an edge $(v,u)$ in $X_k$:
\begin{align*}
x_k(v,u)=\mathbb{I}[(v,u)\in X_k].
\end{align*}
For $k=0$, these are free variables; the remainder are constants (since the control-flow edges $X_i$ are known).
\item {\bf Metadata indicators:} For every $v,v'\in V_k$ (or $V$ if $k=0$) and $d\in\mathcal{D}$, $y_k(v,v',d)$ indicates whether there is a metadata $(v,v',d)\in Y_k$:
\begin{align*}
y_k(v,v',d)=\mathbb{I}[(v,v',d)\in Y_k].
\end{align*}
For $k=0$, these are free variables; the remainder are constants (since the metadata edges $Y_i$ are known).
\end{itemize}

Each assignment to the free variables corresponds to a candidate signature
together with embeddings $F_i$ into the ICCG of each given sample $i$.

\begin{example}
\label{ex:var}
Consider the ICCGs of samples from the GoldDream malware family shown in Figure~\ref{fig:golddream}. Recall that, given these samples, our algorithm uses $V=\{v_1,...,v_8\}$. We first describe the constants constructed by our algorithm, and then the free variables.

The constants include type indicators both for the samples and for the signature, as well as control-flow and metadata indicators for the samples. For the signature, the non-zero type indicators are
\begin{align*}
& t_0(v_\ell,\texttt{\small activity}) =t_0(v_7,\texttt{\small service}) \\
= \ & t_0(v_8,\texttt{\small receiver})=1,
\end{align*}
for $1\le\ell\le6$, and the remaining type indicators are zero.
For the vertex \texttt{zjReceiver} in first sample G1, the type indicators are:
\[
\begin{array}{c}
t_1(\texttt{\small zjReceiver},\texttt{\small receiver})=1
\end{array}
\]
\begin{align*}
& t_1(\texttt{\small zjReceiver},\texttt{\small service})  \\
= \ & t_1(\texttt{\small zjReceiver},\texttt{\small activity})  \\
= \ & 0
\end{align*}

For the outgoing edges from \texttt{zjReceiver}, the control-flow indicators are
\begin{align*}
& x_1(\texttt{\small zjReceiver},\texttt{\small zjService}) \\
= \ & 1
\end{align*}
and
\begin{align*}
&x_1(\texttt{\small zjReceiver},\texttt{\small zjReceiver}) \\
= \ & x_1(\texttt{\small zjReceiver},\texttt{\small ISniper})\\
= \ & x_1(\texttt{\small zjReceiver},\texttt{\small moreGame}) \\
= \ & x_1(\texttt{\small zjReceiver},\texttt{\small HighScore}) \\
= \ & x_1(\texttt{\small zjReceiver},\texttt{\small Profile}) \\
= \ & x_1(\texttt{\small zjReceiver},\texttt{\small UserAct}) \\
= \ &x_1(\texttt{\small zjReceiver},\texttt{\small GameAct})\\
= \ & 0
\end{align*}

Assignments to the free variables correspond to candidate signatures along embeddings $F_1$ and $F_2$ into the samples G1 and G2 respectively.
For example, consider the candidate signature S1, along with the candidate embedding $F_1^{(a)}$ into G1 described in Example~\ref{ex:prob}, and the candidate embedding
\begin{align*}
\begin{cases}
F_2^{(a)}(\texttt{receiver0})=\texttt{MyReceiver} \\
F_2^{(a)}(\texttt{service0})=\texttt{BoolService}
\end{cases}
\end{align*}
into G2. Then, the domain is represented as  $V_0=\{\texttt{service0},\texttt{receiver0}\}$, so the assignments to the domain indicators are
\begin{align*}
&d(v_\ell)\mapsto0\hspace{0.25in}(\text{for }1\le\ell\le6)\\
&d(v_7)\mapsto1,\hspace{0.25in} d(v_8)\mapsto1.
\end{align*}
The assignments to the embedding indicators for $F_1^{(a)}$ are
\begin{align*}
&f_1(v_7,\texttt{zjReceiver})\mapsto1, \ \ f_1(v_8,\texttt{zjService})\mapsto1 \\
&f_1(v_7,w)\mapsto0\hspace{0.25in}(\text{for }w\not=\texttt{zjReceiver}) \\
&f_1(v_8,w)\mapsto0\hspace{0.25in}(\text{for }w\not=\texttt{zjService}) \\
&f_1(v_\ell,w)\mapsto0\hspace{0.25in}(\text{for }1\le\ell\le6\text{ and }w\in V_1)\\
\end{align*}
The assignments to the control-flow indicators are
\begin{align*}
&x_0(v_7,v_8)\mapsto1, \ \ x_0(v_\ell,v_{\ell'})\mapsto0\hspace{0.1in}(\text{for }(\ell,\ell')\not=(7,8)).
\end{align*}

If instead the candidate embedding of the signature S1 into the sample G1 were $F_1^{(b)}$ (also described in Example~\ref{ex:prob}), then the assignments to the embedding indicators would become
\begin{align*}
&f_1(v_7,\texttt{zjReceiver})\mapsto1, \hspace{0.25in} f_1(v_8,\texttt{GameAct})\mapsto1 \\
&f_1(v_7,w)\mapsto0\hspace{0.25in}(\text{for }w\not=\texttt{zjReceiver}) \\
&f_1(v_8,w)\mapsto0\hspace{0.25in}(\text{for }w\not=\texttt{GameAct}) \\
&f_1(v_\ell,w)\mapsto0\hspace{0.25in}(\text{for }1\le\ell\le6\text{ and }w\in V_1),
\end{align*}
and assignments to the remaining free variables would be unchanged.

\end{example}

\subsection{Encoding of Common Subgraph}

We now describe how to encode the requirement that $(V_0, X_0, Y_0)$ should be a common subgraph of all the samples.  Recall from earlier 
that the common subgraph requirement corresponds to our hard constraints.

\begin{itemize}
\item {\bf Constant domain.} For every $v\in V$, the domain of $F_i$ is the same for each $i$:
\begin{align*}
\bigwedge_i\left\{d(v)=\bigvee_{w\in V_i}f_i(v,w)\right\}
\end{align*}
Here, the conjunction over $i$ states that $d(v)$ should be assigned to true if $v$ is in the domain of the embedding $F_i$. Therefore, this constraint ensures that each $v$ is either in the domain of $F_i$ for every $i$, or not in the domain of $F_i$ for any $i$.
\item {\bf Function property.} For every $v\in V$ and $w,w'\in V_i$ where $w\not=w'$, $F_i$ cannot map $v$ to both $w$ and $w'$.
\begin{align*}
(\neg f_i(v,w))\vee(\neg f_i(v,w')).
\end{align*}
\item {\bf One-to-one.} For every $v,v'\in V$ where $v\not=v'$ and $w\in V_i$:
\begin{align*}
(\neg f_i(v,w))\vee(\neg f_i(v',w)).
\end{align*}
\item {\bf Type preserving.} For every $v\in V$, $w\in V_i$, and $r\in\mathcal{T}$, we have:
\begin{align*}
f_i(v,w)\Rightarrow(t_0(v,r)=t_i(w,r)).
\end{align*}
\item {\bf Control-flow preserving.} For every $v,v'\in V$ and $w,w'\in V_i$:
\begin{align*}
f_i(v,w)\wedge f_i(v',w')\wedge x_0(v,v')\Rightarrow x_i(w,w').
\end{align*}
\item {\bf Metadata preserving.} For every $v,v'\in V$ and $d\in\mathcal{D}$, we have:
\begin{align*}
f_i(v,w)\wedge f_i(v,w')\wedge y_0(v,v',d)\Rightarrow y_i(w,w',d).
\end{align*}
\item {\bf No spurious control-flow.} For every $v,v'\in V$, the edges $X_0$ are a subset of $V_0\times V_0$:
\begin{align*}
x_0(v,v') \Rightarrow d(v), \ \ x_0(v,v') \Rightarrow d(v')
\end{align*}
\item {\bf No spurious metadata.} For every $v,v'\in V$ and $d\in\mathcal{D}$, the metadata $Y_0$ are a subset of $V_0\times V_0\times\mathcal{D}$:
\begin{align*}
y_0(v,v',d) \Rightarrow d(v), \ \ 
y_0(v,v',d)&\Rightarrow d(v')
\end{align*}
\end{itemize}

\begin{example}
Consider the ICCGs for the samples G1 and G2 from the GoldDream family shown in Figure~\ref{fig:golddream}, along with the candidate signature S1. In Example~\ref{ex:var} we described the constants in our MaxSAT encoding of the signature synthesis problem, as well as the assignments to free variables corresponding to the candidate signature S1 together with candidate embeddings $F_2^{(a)}$ and either $F_1^{(a)}$ or $F_1^{(b)}$.

Recall that $F_1^{(a)}$ satisfies the properties described in Section~\ref{sec:prob} (as does $F_2^{(a)}$), whereas $F_1^{(b)}$ is neither type preserving nor edge preserving. In particular, the candidate embedding $F_1^{(b)}$ has corresponding free variable assignments
\begin{align*}
&x_0(v_7,v_8)\mapsto1, \  f_1(v_7,\texttt{zjReceiver})\mapsto1,\\
&f_1(v_8,\texttt{GameAct})\mapsto1.
\end{align*}
These assignments violate the type preservation constraint
\begin{align*}
&f_1(v_8,\texttt{GameAct}) \\
&\Rightarrow(t_0(v_8,\texttt{activity})=t_1(\texttt{GameAct},\texttt{activity}))
\end{align*}
because one of the subterms $t_0(v_8,\texttt{activity})=0$ but $t_1(\texttt{GameAct},\texttt{activity})=1$, so $F_1^{(b)}$ does not satisfy the type preservation constraints. In addition, these assignments violate the control-flow preservation constraint
\begin{align*}
&f_1(v_7,\texttt{zjReceiver})\wedge f_1(v_8,\texttt{GameAct})\wedge x_0(v_7,v_8) \\
&\Rightarrow x_1(\texttt{zjReceiver},\texttt{GameAct})
\end{align*}
since the constant $x_1(\texttt{zjReceiver},\texttt{GameAct})=0$, so $F_1^{(b)}$ is not control-flow preserving.

On the other hand, it is not difficult to verify that the candidate signature S1 together with the candidate embeddings $F_1^{(a)}$ and $F_2^{(a)}$ correspond to an assignment that satisfies the constraints described above.
\end{example}

\begin{comment}
We briefly describe how to convert each kind of constraint to a linear (in)equality:
\begin{itemize}
\item Constant domain. These can be expressed as
\begin{align*}
d(v)=\sum_{w\in V_i}f_i(v,w).
\end{align*}
This is because $f_i(v,w)=1$ for at most one $w\in V_i$.
\item Function property. These can be expressed as
\begin{align*}
f_i(v,w)+f_i(v,w')\le1.
\end{align*}
\item One-to-one. These can be expressed as
\begin{align*}
f_i(v,w)+f_i(v',w)\le1.
\end{align*}
\item Type preserving. These can be expressed as
\begin{align*}
f_i(v,w)&\le1+t_0(v,r)-t_i(w,r) \\
f_i(v,w)&\le1-t_0(v,r)+t_i(w,r).
\end{align*}
To see this, note that
\begin{align*}
-1\le t_0(v,r)-t_i(v,r)\le1,
\end{align*}
so the constraints hold trivially unless $f_i(v,w)$ is true.
If $f_i(v,w)$ is true, then the constraints reduce to
\begin{align*}
0&\le t_0(v,r)-t_i(w,r) \\
0&\le -t_0(v,r)+t_i(w,r)
\end{align*}
which holds if and only if $t_0(v,r)=t_i(w,r)$.
\item Control-flow preserving. These can be expressed as
\begin{align*}
f_i(v,w)+f_i(v',w')+x_0(v,v')&\le 2+x_i(w,w').
\end{align*}
\item Metadata preserving. These can be expressed as
\begin{align*}
x_i(v,w)+x_i(v,w')+y_0(v,v',d)\le2+y_i(w,w',d).
\end{align*}
\item No spurious control-flow. These can be expressed as
\begin{align*}
x_0(v,v')&\le d(v) \\
x_0(v,v')&\le d(v').
\end{align*}
\item No spurious metadata. These can be expressed as
\begin{align*}
y_0(v,v',d)&\le d(v) \\
y_0(v,v',d)&\le d(v').
\end{align*}
\end{itemize}
\end{comment}

\subsection{Encoding of Maximally Suspicious}
\label{subsec:function}

Now, we describe how to encode the maximally suspicious requirement, 
which is that the synthesized signature $(V_0,X_0,Y_0)$ has a maximal 
suspiciousness $|X_0|+\sum_{y\in Y_0}w_y$, where the weight $w_y$ indicates the relative importance of  metadata edge $y$ (as described in Section~\ref{sec:prob2}).
In particular, we maximize the objective
\begin{align*}
\mathcal{O}=\sum_{v,v'\in V}x_0(v,v') + \sum_{v,v'\in V}\sum_{d\in\mathcal{D}} w_{(v,v',d)} y_0(v,v',d).
\end{align*}

To choose the weights $w_y$, we use the frequency of the metadata in benign samples to assign weights to different kinds of metadata. In particular, we define $w_y$ to be:
\begin{align*}
w_y=\frac{\#\{\text{benign apps}\}+1}{\#\{\text{benign apps containing }y\}+1}
\end{align*}
In other words, $w_y$ is the  inverse frequency with which  metadata edge $y$ occurs in benign samples, computed on a large corpus of benign apps (we add one to avoid division by zero). The intuition is that metadata edges in the malware samples that rarely occur in benign apps are more likely to correspond to malicious behaviors.

However, some kinds of behaviors in $Y_0$ are strictly more dangerous than others.
In particular, according to previous literature~\cite{drebin,DroidRanger},
suspicious intent filters are far more likely to indicate malicious behavior than API calls, which are in turn much more suspect than data-flows:
\begin{align*}
\text{intent filters }>\text{ API calls }>\text{ data-flows}.
\end{align*}
Rather than simply optimizing the weighted sum, we first prefer signatures that have the highest weighted sum restricted to intent filters, regardless of other metadata.
Ties are broken by the weighted sum restricted to suspicious API calls, and further ties are broken by the weighted sum restricted to data-flows.
The number of edges in $X_0$ is considered last, since it is already indirectly optimized by the other objectives.
To incorporate this strict ordering on different kinds of metadata,
we group the sums in the objective according to the different kinds of metadata
and then encode the objective as a lexicographical optimization problem in the MaxSAT solver~\cite{lexico}.

\section{Approximate Signature Matching}
\label{sec:approximate}

In addition to making semantic malware detection fully automatic, another benefit of our signature inference algorithm is that it enables \emph{approximate signature matching}. Suppose we are given an app $\mathcal A$, and we want to determine if $\mathcal A$ is an instance of malware family $\mathcal F$. Even if $\mathcal A$ does not exactly match $\mathcal F$'s signature, we might want to determine if $\mathcal A$ exhibits a high degree of similarity with other instances of $\mathcal F$. This problem, which we refer to as \emph{approximate matching}, is useful both for detecting zero-day malware and also for mitigating  behavioral obfuscation. This section explains how we  leverage signature inference for approximate signature matching.

\begin{comment}
Given a signature $S$ and an app $\mathcal A$, \apposcopy checks if $\mathcal A$ contains $S$, i.e. it checks if $S$ is an isomorphic subgraph of $\mathcal A$. Even though \apposcopy is resilient to low-level obfuscation techniques (such as renaming methods), it can be evaded by changing the structure of the ICCG. To mitigate this problem \toolname uses an approximate signature matching that is not only resilient to low-level obfuscation techniques but also to behavioral obfuscation. 

The matching algorithm of \toolname works as follows. First, we compute the transitive closure $\tau_{\mathcal A}$ of the ICGG of $\mathcal A$. For every pair of nodes $n_i$, $n_j$ in the ICCG, if there exists a path from $n_i$ to $n_j$, then we add an edge from $n_i$ to $n_j$ to the ICCG. Suppose an attacker obfuscates $\mathcal A$ by randomly introducing nodes between $n_i$ and $n_j$. This obfuscation would break the common isomorphic subgraphs between the ICCG of $\mathcal A$ and $S_1$ and would avoid the detection of \toolname. However, by computing $\tau_{\mathcal A}$, \toolname is now resilient to this kind of obfuscation.  
\end{comment}

To perform approximate matching between an app $\mathcal A$ and a malware family $\mathcal F$ with signature  $\mathcal{S_F}$, we first assume that $\mathcal A$ and $\mathcal{S_F}$ belong to the same malware family. We then compute a new signature $\mathcal S$ that captures the common behavior of  $\mathcal A$ and all instances of $\mathcal F$.
If  $\mathcal{S_F}$ and $\mathcal S$ are ``similar'' (see below), there is a high probability that $\mathcal A$ is an obfuscated instance of malware family $\mathcal F$. 

To understand how we measure similarity, note that $\mathcal S$ is always a subgraph of  $\mathcal{S_F}$; hence, we can measure similarity in terms of number of nodes and edges that are removed from  $\mathcal{S_F}$ to form $\mathcal S$. Specifically, given signature $\mathcal S$, let $f({\mathcal S})$ be a function that measures the size of $S$ as a weighted sum of the number of nodes and edges in $\mathcal S$. Then, given app $\mathcal A$ and family $\mathcal F$ with signature  $\mathcal{S_F}$, we  define the similarity metric as follows:

\begin{equation*}
\delta(\mathcal A,  \mathcal{F}) = \frac{f({\textsc{InferSignature}(\mathcal A, \mathcal{S_F})})}{f(\mathcal{S_F})}
\end{equation*}
Hence, if $\delta(\mathcal A,  \mathcal{F})$ is sufficiently high, then $\mathcal A$ is more likely to be an instance  of family $\mathcal F$. As we show in Section~\ref{sec:eval}, approximate matching using this similarity metric between signatures makes \toolname more resilient to behavioral obfuscation.

\vspace{0.05in}
\noindent
{\bf \emph{Zero-day malware detection.}} We can also use  approximate signature matching  to detect zero-day malware.
Suppose  we have a database of signatures 
for existing malware families $\mathcal F_1, \ldots, \mathcal F_n$, and suppose that an app $\mathcal A$ does not match any of them. Now, to determine whether $\mathcal A$ belongs to a new (unknown) malware family, we  compute $\delta(\mathcal A, \mathcal F_i)$ for each $ 1 \leq i \leq n$ and report $\mathcal A$ as malware if $\delta(\mathcal A, \mathcal F_i)$ exceeds a certain cutoff value for \emph{some} malware family $\mathcal F_i$.
We explain how we pick this cutoff value in Section~\ref{sec:impl}.

\section{Implementation}
\label{sec:impl}

We have implemented the proposed technique in a
tool called \toolname \ for inferring semantic malware signatures
for Android.
Our implementation consists of about 7,000 lines of Java code
and uses the \wbo MaxSAT solver~\cite{openwbo}. Our implementation 
builds on top of \apposcopy~\cite{apposcopy} for statically constructing ICCGs and taint flows of 
Android apps. Specifically, \apposcopy implements
a field- and context-sensitive pointer analysis
and constructs a precise callgraph  by 
simultaneously refining the targets of virtual method calls and points-to sets. 
For context-sensitivity, \apposcopy uses the hybrid approach proposed in~\cite{hybrid13}. Specifically, it uses call-site sensitivity for static method
calls and object-sensitivity for virtual method calls. 
To scale in large apps, \apposcopy also leverages the \textsc{Explorer}~\cite{explorer} tool to construct ICCGs in a demand-driven manner.
\apposcopy's average static analysis time for constructing an ICCG is 87 seconds for an app from the Android Genome benchmarks and 126 
seconds for an app from Google Play, including analysis time for all third-party libraries. Unlike DroidSafe~\cite{droidsafe} which analyzes the source code of Android framework directly, \apposcopy uses about 1,210 manually-written models for classes that are relevant to its underlying taint analysis.

Once a signature for a given malware family is generated, \toolname can perform both exact matching (described in Section~\ref{sec:prob}) as well as approximate matching (described in Section~\ref{sec:approximate}). Our implementation of the approximate matching algorithm differs slightly from the description in Section~\ref{sec:approximate}; in particular, we use the transitive closure of the control-flow edges in the ICCG (computed in a preprocessing step) rather than the original ICCG control-flow edges; this modification increases the resilience of \toolname against obfuscations that introduce dummy components.

We use approximate matching both to detect obfuscated apps and to detect zero-day malware. Recall from Section~\ref{sec:approximate} that our approximate matching algorithm uses a cutoff value for how high the similarity metric must to count as a match. We choose a cutoff of 0.5 for zero-day malware, and a stricter cutoff of 0.8 for obfuscated malware. In other words, an app with a similarity score $=1.0$ is flagged as an unobfuscated instance of a known malware family, an app with similarity score $>0.8$ is flagged as an obfuscated instance of a known malware family, and an app with similarity score $>0.5$ is flagged as a possible zero-day malware.

We describe our methodology for choosing the similarity metric cutoff for zero-day malware detection; the cutoff for obfuscated malware is chosen similarly. We use the Android Malware Genome Project as a training set. For each family $\mathcal F$ in the Android Malware Genome Project, we omit the signature for $\mathcal F$ from \toolname's database of signatures (so samples in $\mathcal F$ appear as zero-day malware to \toolname). Then, we use the remaining signatures in the database to try and detect samples in $\mathcal F$ using our zero-day malware detection algorithm in Section~\ref{sec:approximate}, using each cutoff value 0.6382,  0.5834, 0.5832, 0.4927, and 0.4505 (selected by manually binning the data).

For each cutoff value and each $\mathcal F$, we compute the true positive rate (i.e., the fraction of samples in $\mathcal F$ that \toolname detects); then, for each cutoff value, we take the weighted average over families $\mathcal F$ to obtain an overall true positive rate for that cutoff value. Finally, we select the largest possible cutoff that still achieves a 90\% true positive rate. The selected cutoff is 0.4927, which we round to 0.5. Figure~\ref{fig:zero} shows the ROC curve obtained using the various cutoff values, with the $y$-axis showing the computed true positive rate. The $x$-axis shows the corresponding false positive rate, which is computed on an independent test set of benign apps.

\definecolor{bblue}{HTML}{188FA7}
\usetikzlibrary{pgfplots.patchplots}

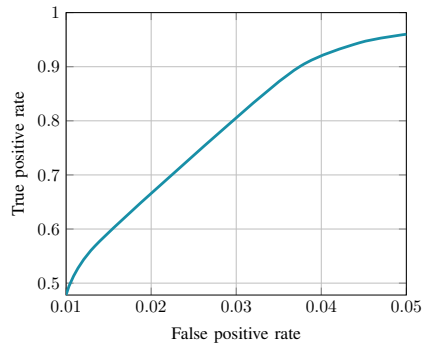
\begin{figure}[!t]
\centering
\scalebox{0.66}{
\begin{tikzpicture}
\begin{axis}[
  ymajorgrids = true,
  xmajorgrids=true,
  enlarge x limits=0,
  enlarge y limits=0,
  xtick={0,0.01,0.02,0.03,0.04,0.05},
  ytick={0,0.1,0.2,0.3,0.4,0.5,0.6,0.7,0.8,0.9,1.0},
  xmin=0.01,
  xmax=0.05,
  ymax=1.0,
  scaled ticks=false,
  tick label style={/pgf/number format/fixed},
  ylabel = {True positive rate},
  ylabel style={yshift=-3mm},
  xlabel = {False positive rate}
  ]
\addplot[smooth,ultra thick,bblue]
coordinates {
(0.01,0.478) (0.0134,0.5687) (0.0296,0.8) (0.0376,0.9015) (0.0445,0.9443) (0.05,0.96)
};
\end{axis}
\end{tikzpicture}
}
\caption{Effect of varying cutoff values on false positives and false negatives for zero-day malware detection}
\label{fig:zero}
\vspace*{-5mm}
\end{figure}

\definecolor{bblue}{HTML}{0064FF}
\definecolor{rred}{HTML}{C0504D}
\definecolor{ggreen}{HTML}{9BBB59}
\definecolor{ppurple}{HTML}{9F4C7C}

\begin{figure*}[!t]
\centering
\begin{subfigure}{.6\textwidth}
\centering
\scalebox{0.63}{
\begin{tikzpicture}
    \begin{axis}[
        width  = 8cm,
        height = 8cm,
        y=0.1cm,
        x=1.2cm,
        major x tick style = transparent,
        ybar=2*\pgflinewidth,
        bar width=11pt,
        ymajorgrids = true,
        ylabel = {Accuracy},
        ylabel style={yshift=-4mm},
        symbolic x coords={A,B,C,D,E,F,G,H,I,J,K,L,M},
        xtick = data,
        scaled y ticks = false,
        ymin=0,
        ymax=100,
        legend cell align=left,
        legend entries={Manual Signature, \toolname, Avg. Accuracy Manual Signature, Avg. Accuracy \toolname},
        legend style={
                at={(0.5,1.08)},
                legend columns=-1,
                anchor=north,
        }
    ]
        \addplot[style={ggreen,fill=ggreen,mark=none}]
         coordinates{(A,96.6)(B,98.9)(C,38.0)(D,97.1)(E,100.0)(F,97.8)(G,83.7)(H,100.0)(I,100.0)(J,92.9)(K,100.0)(L,100.0)(M,100.0)};

        \addplot[style={ppurple,fill=ppurple,mark=none}]
         coordinates{(A,94.1)(B,100.0)(C,70.8)(D,100.0)(E,97.8)(F,97.8)(G,100.0)(H,100.0)(I,100.0)(J,93.8)(K,100.0)(L,100.0)(M,100.0)};

\coordinate (A) at (axis cs:A,93.8);
\coordinate (B) at (axis cs:A,90.0);
\coordinate (O1) at (rel axis cs:0,0);
\coordinate (O2) at (rel axis cs:1,0);

\draw [ggreen,sharp plot,dashed,line width=0.8mm] (B -| O1) -- (B -| O2);         
\draw [ppurple,sharp plot,dashed,line width=0.8mm] (A -| O1) -- (A -| O2);

\addlegendimage{line legend,ggreen,sharp plot,dashed,line width=0.8mm}
\addlegendimage{line legend,ppurple,sharp plot,dashed,line width=0.8mm}

    \end{axis}
\end{tikzpicture}
}
\end{subfigure}
\begin{subfigure}{.35\textwidth}
\centering
\hspace*{5mm}
\scalebox{0.9}{
\begin{tabular}{llc}
\hline
Id & Malware Family & \#Samples\\
\hline
A & DroidKungFu & 444\\
B & AnserverBot & 184\\
C & BaseBridge & 121\\
D & Geinimi & 68\\
E & DroidDreamLight & 46\\
F & GoldDream & 46\\
G & Pjapps & 43\\
H & ADRD & 22\\
I & jSMSHider & 16\\
J & DroidDream & 14\\
K & Bgserv & 9\\
L & BeanBot & 8\\
M & GingerMaster & 4\\
\hline
\end{tabular}
}
\end{subfigure}
\caption{Accuracy of \apposcopy\xspace with a manual and an automated signature synthesized by \toolname}
\label{fig:manual-signature}
\end{figure*}

\section{Evaluation}\label{sec:eval}

The goal of our evaluation is to answer the following questions:

\begin{itemize}
\item[{\bf Q1.}] How do the signatures synthesized by \toolname compare with manually-written signatures used for evaluating \apposcopy in terms of precision and accuracy?
\item[{\bf Q2.}] How do the quality of learned signatures improve as we increase the number of samples?
\item[{\bf Q3.}] How does \toolname compare against other state-of-the-art malware detectors?
\item[{\bf Q4.}] How effective is \toolname at detecting zero-day malware?
\item[{\bf Q5.}] How resistant is \toolname to behavioral obfuscation?
\end{itemize}

In what follows, we describe a series of five experiments designed to answer these questions. All experiments are conducted on an Intel Xeon(R) computer with an E5-1620 v3 CPU and 32G of memory running on Ubuntu 14.04.

\subsection{Astroid vs. Manual Signatures}
\label{subsec:evalmanual}

In our first experiment, we compare the signatures synthesized by \toolname with manually written signatures used for evaluating \apposcopy~\cite{apposcopy}. 
Since \toolname generates semantic malware signatures in the specification language used in \apposcopy, we use \apposcopy's (exact) signature matching algorithm to decide if an app matches a signature.
According a co-author of~\cite{apposcopy}, it took \emph{several weeks} to manually construct signatures for all malware families in their dataset, even with full knowledge of the malware family of each app and the nature of the malice.

\vspace{0.05in}
\noindent {\bf \emph{Accuracy on known malware.}} Since the authors of~\cite{apposcopy} have manually written semantic signatures for all malware families from the Android Malware Genome Project~\cite{malware-genome}, we use \toolname to synthesize signatures for malware families from this dataset.
The table in Figure~\ref{fig:manual-signature} shows the malware family names and the number of samples for each family. For each family $\mathcal{F}$, we randomly select 5 samples from $\mathcal{F}$ and use \toolname to infer a signature for $\mathcal{F}$ from these samples. Since different sets of samples may produce different signatures, we run \toolname with 11 different sets of randomly chosen samples, and report our results for the signature that achieves the median accuracy out of these 11 runs. 

The plot in Figure~\ref{fig:manual-signature} compares the accuracy \toolname against \apposcopy. Here, accuracy is the number of correctly classified samples divided by the total number of samples. For most malware families,  \toolname yields similar or better accuracy than \apposcopy, and the overall accuracy of \toolname (93.8\%) is higher than that of \apposcopy (90\%). Furthermore, for two malware families (specifically, Pjapps and BaseBridge), \toolname generates significantly better signatures than manually written ones. 
In summary, these results show that \toolname can achieve \emph{significantly fewer false negatives} compared to \apposcopy.

\vspace{0.05in}
\noindent {\bf \emph{False positives on known malware.}}
\toolname reports zero  false positives on the Android Malware Genome Project samples. In contrast, \apposcopy reports two such false positives (specifically, it misclassifies two instances of other malware families as belonging to Geinimi). Hence, the signatures synthesized by  \toolname outperform manually written ones, both in terms of accuracy as well as precision.

\vspace{0.05in}
\noindent {\bf \emph{False positives on benign apps.}}
 To determine whether \toolname produces false positives on benign apps, we analyze a corpus of 10,495 apps downloaded from Google Play during 2013-14. According to~\cite{DroidRanger},  over $99.9\%$ of these apps are known to be benign.  \toolname reports a total of 8 malicious apps (specifically, 2 instances each of DroidDream and DroidDreamLight, 1 instance of Pjapps, and 3 instances of DroidKungFu). We uploaded these 8 apps to VirusTotal~\cite{virus-total},
which is a service that provides aggregated reports from multiple anti-virus tools. VirusTotal agreed with \toolname on each of these 8 apps.\footnote{To account for the inclusion of some lower-quality AV tools in VirusTotal results, we only consider VirusTotal to report an app as malware if the majority (i.e., more than half) of AV tools classify the app 
to be malicious.} 
In other words, \toolname reported \emph{zero false positives} of either kind.

\definecolor{bblue}{HTML}{0064FF}
\definecolor{rred}{HTML}{C0504D}
\definecolor{ggreen}{HTML}{9BBB59}
\definecolor{ppurple}{HTML}{9F4C7C}

\begin{figure}[!t]
\centering
\scalebox{0.85}{
\begin{tikzpicture}
\pgfplotsset{
    scale only axis
}

    \begin{axis}[
        axis y line*=left,
        ybar=2*\pgflinewidth,
        bar width=11pt,
        ymajorgrids = true,
        ylabel = Accuracy,
        xlabel = Number of samples,
        ylabel style={yshift=-3mm},
        symbolic x coords={2,3,4,5,6,7,8,9,10},
        xtick={2,3,4,5,6,7,8,9,10},
        ytick={0,10,20,30,40,50,60,70,80,90,100},
        ymin=0,
        ymax=100,
        y=0.04cm,
        legend entries={Accuracy, Time},
        legend style={
                at={(0.5,1.2)},
                legend columns=-1,
                anchor=north,
        }
    ]
    
        \addplot[style={ppurple,fill=ppurple,mark=none}]
         coordinates{(2,90.8)(3,93.0)(4,93.3)(5,93.8)(6,93.8)(7,93.8)(8,93.8)(9,93.8)(10,93.8)};

\addlegendimage{line legend,bblue,smooth,line width=0.5mm}
    \end{axis}

    \begin{axis}[
        ylabel near ticks, yticklabel pos=right,
        ylabel style={yshift=1mm},
        axis x line=none,
        ylabel=Seconds,
        ymin=0,
        ymax=5,
        y=0.8cm,
        ytick={0,1,2,3,4,5},
                legend style={
                at={(0.8,1.3)},
                legend columns=-1,
                anchor=north,
        }
        ]
    \addplot[smooth,mark=*,bblue,line width=0.5mm] plot coordinates {
        (2, 1.41)
        (3, 1.45)
        (4, 1.47)
        (5, 1.51)
        (6, 1.52)
        (7, 1.52)
        (8, 1.52)
        (9, 1.75)
        (10, 1.81)
    };

    \end{axis}
    \end{tikzpicture}
    }
    \caption{Accuracy and time to synthesize the signature with different number of samples}
    \label{fig:sample-accuracy}
\end{figure}
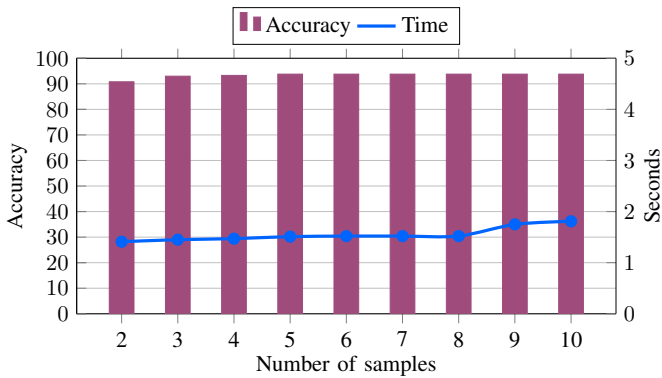
\subsection{Varying the Number of Samples}

We evaluate the performance and accuracy of \toolname with respect to the number of samples. In Figure~\ref{fig:sample-accuracy}, the blue line plots  signature inference time (in seconds) against the number of samples. \toolname is quite fast, taking less than two seconds on average to infer a signature for a malware family. Since manually writing malware signatures typically requires \emph{several hours} of human effort, we believe these statistics show that \toolname is quite practical. 

The bars in Figure~\ref{fig:sample-accuracy} show how the accuracy varies with respect to the number of samples. \toolname achieves 90.8\% accuracy using \emph{only two samples}. When we increase the number of samples to 5, \toolname achieves 93.8\% accuracy. More samples do not seem to benefit, likely due to imprecision in the static analysis for ICCG constructions. These results show that \toolname can infer very good signatures from a very small number of malware samples.

\subsection{Comparison Against Existing Tools}

We compare \toolname to two state-of-the-art malware detection tools, namely \drebin ~\cite{drebin} (a learning-based malware detector) and \massvet~\cite{MassVet} (which detects repackaged malware). We do not explicitly compare \toolname against anti-virus tools (e.g., VirusTotal) because \toolname is an extension of \apposcopy, which has already been shown to outperform leading anti-virus tools in the presence of obfuscated malware. 

\vspace{0.05in}
\noindent {\bf \emph{Comparison to Drebin.}}
We first compare \toolname against \drebin~\cite{drebin}, a state-of-the-art malware detector based on machine learning. Since our goal is to infer malware signatures using few samples, we   focus on the setting where only five samples are available for the target  family.  
We believe this setting is very important because many malware families have very few known samples. For example, in the Drebin dataset, only 42\% of families have more than 5 samples. 

\drebin takes as input a set of apps and extracts features  using static analysis. It then trains an SVM on these feature vectors to classify apps as malicious or benign. Unlike \toolname, which classifies each app into a specific malware family, \drebin only determines whether an  app is malicious or benign. Since the implementation of \drebin is unavailable, we used their publicly available feature vectors and train SVMs based on the methodology described in~\cite{drebin} using the libsvm library~\cite{CC01a}. For each malware family $\mathcal{F}$, we start with the full \drebin training set, but remove $\mathcal{F}$ except for 5 randomly chosen samples from $\mathcal{F}$ (to achieve our small-sample setting). In summary,  we train \drebin on a large number of benign apps, \emph{all} samples from other malware families, and 5 samples from the target family.  As with \toolname, we train the SVM with 11 different sets of randomly chosen samples and report the median accuracy.

In order to perform an apples-to-apples comparison, we use the 1,005 samples in the Android Malware Genome Pro\-ject for which \drebin feature vectors are available.\footnote{In this experiment, we use the Android Malware Genome Project dataset as opposed to the \drebin dataset because some of the malware families in the \drebin dataset are mislabeled, which is problematic for multi-label classification (as done by our technique). For example, some of the family labels in the Drebin dataset do not match the labels  in the Malware Genome project.}
 We use two cutoffs to evaluate the accuracy on these apps. The first achieves a false positive rate of 1\%, which is used in the original evaluation~\cite{drebin}. However, 1\% false positives still amounts to more than 10,000 false positives on Google Play, which contains more than a million apps. For a closer comparison to \toolname (which reports zero false positives on Google Play apps), we choose a second cutoff that achieves a false positive rate of 0.01\%.

As shown in Figure~\ref{fig:drebin}, \toolname outperforms \drebin in our small-sample setting: While \toolname achieves 93.8\% accuracy, \drebin yields 70.7\% and 88.9\% accuracy using the 0.01\% and 1\% cutoff values respectively. We believe these results indicate that \toolname compares favorable with existing learning-based detectors in the small-sample setting.

\vspace{0.05in}
\noindent {\bf \emph{Comparison to MassVet.}}
We also compare \toolname against \massvet~\cite{MassVet}, a state-of-the-art tool for detecting repackaged malware---i.e., malware produced by adding malicious components to an existing benign app. \massvet maintains a very large database of existing  apps and checks whether a given app is a repackaged version of an existing one in the database. The authors of \massvet provide a public web service~\cite{MassVet-tool}, which we use to evaluate \massvet.

\definecolor{bblue}{HTML}{0064FF}
\definecolor{rred}{HTML}{C0504D}
\definecolor{ggreen}{HTML}{9BBB59}
\definecolor{ppurple}{HTML}{9F4C7C}

\definecolor{b1}{HTML}{BEE9E8}
\definecolor{b2}{HTML}{188FA7}

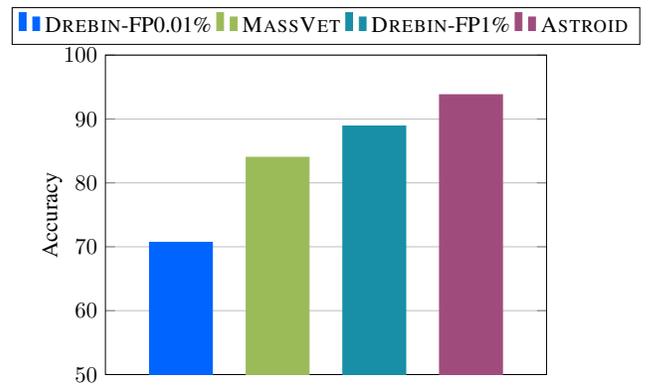
\begin{figure}[!t]
\centering
\scalebox{0.85}{
\begin{tikzpicture}
    \begin{axis}[
        width  = 8cm,
        height = 8cm,
        y=0.1cm,
        x=0.5cm,
        major x tick style = transparent,
        ybar=2*\pgflinewidth,
        bar width=28pt,
        ymajorgrids = true,
        ylabel = {Accuracy},
        ylabel style={yshift=-4mm},
        symbolic x coords={1,2,3,4},
        ytick={50,60,70,80,90,100},
        xticklabels={},
        scaled y ticks = false,
        enlarge x limits= 1.8,
        ymin=50,
        ymax=100,
        legend cell align=left,
        legend entries={\drebin-FP0.01\%, \massvet, \drebin-FP1\%, \toolname},
        legend style={
                at={(0.5,1.15)},
                legend columns=-1,
                anchor=north,
        }
    ]
        \addplot[style={bblue,fill=bblue,mark=none}]
        coordinates{(1,70.7)};

        \addplot[style={ggreen,fill=ggreen,mark=none}]
        coordinates{(2,84.0)};

        \addplot[style={b2,fill=b2,mark=none}]
        coordinates{(3,88.9)};

        \addplot[style={ppurple,fill=ppurple,mark=none}]
        coordinates{(4,93.8)};

    \end{axis}
\end{tikzpicture}
}
\caption{Comparison between \drebin with FP0.01\% and FP1\%, \massvet and \toolname}
\label{fig:drebin}
\end{figure}

As shown in Figure~\ref{fig:drebin}, the overall  accuracy of \massvet on the Android Malware Genome Project is 84.0\%, which is significantly lower than the 93.8\% accuracy of \toolname. 
These false negatives occur because many malware samples are not repackaged versions of existing benign apps, and even repackaged malware may go undetected if the original app is missing from their database.

We also evaluate the false positive rate of \massvet  on a corpus of 503 benign apps from  Google Play.~\footnote{We were unable to apply \massvet to all 10,495 apps from Google Play due to limitations of the web service.}  \massvet reports that 176 of these apps are malware, but all except one of these apps are classified as benign by VirusTotal.
 Since several benign applications repackage existing apps by adding ad libraries, \massvet seems to mistakenly classify them as malware. In summary, this experiment demonstrates that \toolname achieves both higher accuracy as well as a lower false-positive rate compared to \massvet.

\vspace{0.05in}
\noindent {\bf \emph{Interpretability.}} In addition to comparing favorably with \massvet and \drebin in terms of accuracy and false positives, we believe that \toolname produces better explanations of malice compared to existing tools. In the Appendix, we compare the semantic signatures produced by \toolname with the evidence of malice produced by \drebin and \massvet. Since \toolname can pinpoint the malicious components and their suspicious metadata (e.g., sensitive information leaked by the component), we believe that the signatures generated by \toolname are more interpretable (and therefore more helpful) to security analysts.

\subsection{Detection of Zero-day Malware}\label{sec:zero}

To evaluate whether \toolname can effectively detect zero-day malware, we conduct experiments on two different datasets and evaluate \toolname's accuracy and false positive rate.

\vspace{0.05in}
\noindent
{\bf \emph{Malware from Symantec and McAfee.}} In our first experiment, we use 160 malware samples obtained from Symantec and McAfee, two leading anti-virus companies. Even though these applications are known to be malicious, none of them belong to the  malware families  from the Android Malware Genome Project dataset. Since \toolname's database only contains signatures for the families shown in Figure~\ref{fig:manual-signature}, all of these 160 applications are zero-day malware with respect to \toolname's signature database. Using \toolname's approximate signature matching algorithm (with the cutoff of 0.5, as described in Section~\ref{sec:impl}),
\toolname correctly identifies 147 of these apps as malware; hence, \toolname's accuracy for zero-day malware detection on this dataset is 92\%. In contrast, \massvet's accuracy on this dataset is  81\%. We were not able to compare against \drebin on this dataset because we do not have their feature vectors available.

\begin{comment}
three sets of experiments on three different datasets, one on the dataset from the Android Malware Genome Project and the second on apps collected from Google Play.

For the first experiment on the Android Malware Genome dataset,  we pretend that a given malware family $\mathcal F$ is previously unknown by removing its signature from \toolname’s signature database. We then use \toolname’s  approximate signature matching algorithm to check whether some  $\mathcal A \in \mathcal F$ is malware. Figure~\ref{fig:zero} shows a ROC curve where we use different cutoffs~\footnote{The chosen cutoffs are: 0.6382,  0.5834, 0.5832, 0.4927, and 0.4505, which corresponds to false positive rates from 1\% to 5\%, respectively.} for the  similarity metric such that the ratio of false positive for benign apps ranges from 
1\% to 5\%. 
\end{comment}

\vspace{0.05in}
\noindent
{\bf \emph{Apps from Google Play.}}
For our second experiment, we analyze 10,495 Google Play apps using \toolname's approximate signature matching algorithm. As before, \toolname's signature database only contains the malware families from Figure~\ref{fig:manual-signature}. Among these apps, \toolname reports 395 of them (i.e., 3.8\%) as being malicious.
Of these 395 apps, 8 exactly match one of the signatures in \toolname's database; as discussed in Section~\ref{subsec:evalmanual}, these 8 apps are all malicious.

To investigate which of the remaining 387 apps are known malware, we  use VirusTotal to analyze each of these apps. Among these  387 samples, 21 of them are reported as malicious by  the majority (i.e., more than half) of the anti-virus tools, and 81 of them are reported as malicious by at least one anti-virus tool. Of the remaining 306 apps reported by \toolname, we randomly selected 40 apps and manually inspected them. Our manual inspection shows that 22 of these 40 apps are actually malicious since they contain highly suspicious behaviors:

\begin{itemize}
\item 4 apps appear to be SMS Trojans because they automatically block all incoming SMS events when they receive an SMS at a specific time or the contents of the SMS match a certain pattern. 
\item 1 app silently records audio in a background process and aborts all incoming SMS events.
\item 1 app automatically locks the screen and prevents the user from unlocking it.
\item 11 apps dynamically install apk files from untrusted sources and leak the user's device id, IMSI number, and other confidential phone information to untrusted remote servers. 
\item 1 app silently takes pictures (i.e., without the user triggering it) and saves the picture to an encrypted file.
\item 2 apps contain Android.Cauly library, which has recently been classified as PUA (``Potentially Unwanted App")  by Symantec.
\item 2 apps perform highly suspicious actions without the user triggering them. For instance, they  send email or SMS messages to a fixed address or phone number.

\end{itemize}

Hence, our manual inspection shows that \toolname can detect malicious apps that are not identified by existing anti-malware tools. Furthermore, based on our manual inspection, we estimate that of the 306 apps not identified as malicious by existing tools, 55\% are in fact malicious. By this estimate, \toolname's false positive rate is approximately 1.3\% for zero-day malware detection (i.e., using approximate matching with a cutoff of 0.5).

Finally, we remark that the partial match (i.e., the subgraph $\textsc{InferSignature}(\mathcal A, \mathcal{S_F})$ of the ICCG of $\mathcal{A}$) computed by our approximate matching algorithm was indispensable for finding the malicious behaviors in the 40 randomly selected apps. In every case, some part of malicious code was contained in the partial match, allowing us to quickly identify the malicious behavior. In particular, examining all 40 apps took a single analyst only a few hours. As discussed earlier, the interpretability of the inferred signatures is an important feature of \toolname.

\begin{comment}

\definecolor{bblue}{HTML}{188FA7}
\usetikzlibrary{pgfplots.patchplots}

\begin{figure}[!t]
\centering
\scalebox{0.66}{
\begin{tikzpicture}
\begin{axis}[
  ymajorgrids = true,
  xmajorgrids=true,
  enlarge x limits=0,
  enlarge y limits=0,
  xtick={0,0.01,0.02,0.03,0.04,0.05},
  ytick={0,0.1,0.2,0.3,0.4,0.5,0.6,0.7,0.8,0.9,1.0},
  xmin=0.01,
  xmax=0.05,
  ymax=1.0,
  scaled ticks=false,
  tick label style={/pgf/number format/fixed},
  ylabel = {True positive rate},
  ylabel style={yshift=-3mm},
  xlabel = {False positive rate}
  ]
\addplot[smooth,ultra thick,bblue]
coordinates {
(0.01,0.478) (0.0134,0.5687) (0.0296,0.8) (0.0376,0.9015) (0.0445,0.9443) (0.05,0.96)
};
\end{axis}
\end{tikzpicture}
}
\caption{Detection performance of zero-day malware}
\label{fig:zero}
\vspace*{-5mm}
\end{figure}
\end{comment}

\subsection{Detection of Obfuscated Apps}

\definecolor{bblue}{HTML}{0064FF}
\definecolor{rred}{HTML}{C0504D}
\definecolor{ggreen}{HTML}{9BBB59}
\definecolor{ppurple}{HTML}{9F4C7C}

\definecolor{b1}{HTML}{BEE9E8}
\definecolor{b2}{HTML}{188FA7}

\begin{figure}[!t]
\centering
\scalebox{0.85}{
\begin{tikzpicture}
    \begin{axis}[
        width  = 8cm,
        height = 8cm,
        y=0.08cm,
        x=2.8cm,
        major x tick style = transparent,
        ybar=2*\pgflinewidth,
        bar width=24pt,
        ymajorgrids = true,
        ylabel = {Accuracy},
        ylabel style={yshift=-4mm},
        symbolic x coords={Obfuscated,Non-obfuscated},
        ytick={30,40,50,60,70,80,90,100},
        xtick = data,
        scaled y ticks = false,
        enlarge x limits=0.6,
        ymin=30,
        ymax=100,
        legend cell align=left,
        legend entries={Exact, Approximate},
        legend style={
                at={(0.5,1.14)},
                legend columns=-1,
                anchor=north,
        }
    ]
        \addplot[style={ggreen,fill=ggreen,mark=none}]
        coordinates{(Obfuscated,61.0)(Non-obfuscated,93.8)};

        \addplot[style={ppurple,fill=ppurple,mark=none}]
        coordinates{(Obfuscated,94.3)(Non-obfuscated,94.3)};

    \end{axis}
\end{tikzpicture}
}
\caption{Evaluation of different matching techniques}
\label{fig:obs}
\end{figure}
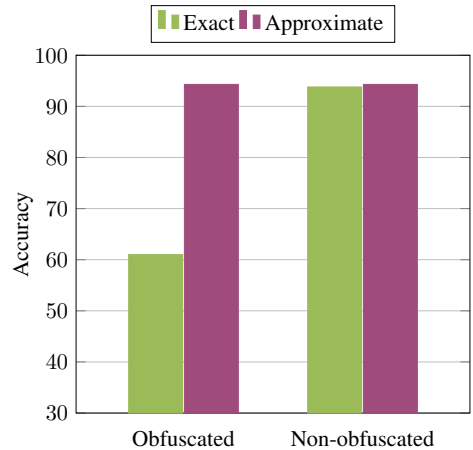

To evaluate whether \toolname is resilient to obfuscations,  we perform a combination of low-level (syntactic) and high-level (semantic) obfuscations. 
First, we obfuscate existing
malware using the ProGuard tool~\cite{proguard}
to rename method/class names, encrypt strings, and modify the program's control flow. Second, 
we also perform obfuscations at the ICCG level, such as inserting dummy components and removing taint flows.

Figure~\ref{fig:obs} shows the accuracy of signature matching using exact vs. approximate matching  for 1,025 malware samples from the Android Malware Genome Project and their corresponding obfuscated versions.   When using the exact matching algorithm of \apposcopy, \toolname can detect 93.8\% of the non-obfuscated malware samples but only 61\% of the obfuscated malware samples.
In contrast, when we use the approximate matching algorithm (with the cutoff of 0.8, as described in Section~\ref{sec:impl}), \toolname is able to detect 94.3\% of malware samples for both obfuscated and non-obfuscated versions. Hence, this experiment demonstrates that \toolname's approximate signature matching algorithm significantly increases the resilience of \toolname to behavioral obfuscations. In contrast, other anti-virus tools (e.g., Symantec, McAfee, Kaspersky etc.) have been shown to be much less resilient to even a subset of the obfuscations that we employ in this experiment~\cite{apposcopy}.

\vspace{0.05in}
\noindent {\bf \emph{False positives.}}
Both exact and approximate matching report zero false positives on malware samples. As before, with exact matching, \toolname reports that 8 apps are malicious from the corpus of 10,495 Google Play apps, but all of these apps are classified as malware by VirusTotal.  When using approximate matching, \toolname reports a total of 13 apps as malicious, 9 of which are classified as malware by VirusTotal. Hence, the false positive rate of \toolname remains very low ($<$0.04\%) even with approximate matching.

\section{Limitations}

Like any other malware detection tool, \toolname \ has a number of limitations:

First, the quality of the signatures inferred by \toolname \ depends on the precision of the underlying static analysis used to construct the ICCG of the samples. In particular, sources of imprecision (or unsoundness) in the analysis can degrade the quality of the signatures inferred by \toolname. 
\begin{comment}
\todo{For instance, \toolname handles reflective calls of which the arguments are unencrypted strings and it considers 
a finite number of models for native methods adopted from \textsc{chord}~\cite{jchord}.}
\end{comment}
However, our experiments indicate that \toolname\ can synthesize high-quality signatures despite possible imprecision in the static analysis. 

Second, \toolname's signature matching algorithm (both the exact and approximate variants) are also affected by the quality of the underlying static analysis. Since signature matching requires computing the ICCG of the application under analysis, any source of unsoundness in the analysis may translate into false negatives in the context of malware detection. For example, if an app dynamically loads a malicious payload, then \toolname \ may fail to flag it as malware. However, such attempts to escape detection can be identified as suspicious, thereby requiring further scrutiny.
On the other hand, sources of imprecision due to the underlying static analysis may also translate into false positives. For 
instance, given a benign app, if the analysis generates a lot of spurious taint flows and inter-component call edges, 
\toolname may mistakenly mark it as malware. However, our evaluation show that the underlying static analysis achieves a high precision without sacrificing scalability.

Third, \toolname requires an analyst to provide at least two representative samples of a given malware family, so  there is still a minimal amount of human effort involved in using \toolname. However,  we believe this effort is miniscule compared to the laborious task of  writing malware signatures manually. 

Finally, \toolname's accuracy in detecting malware is dependent on the initial database of malware signatures. The larger the database, the higher the accuracy in detecting  existing and zero-day malware. However, our experiments show that \toolname can effectively detect new malware families even though we added only thirteen malware signatures to its database. Furthermore, we believe that it is easier to maintain a database of  malware signatures compared to the task of maintaining a much larger database of individual malware samples.

\section{Related Work}

Android malware detection and classification have been extensively
studied in recent years. In this section we briefly discuss prior closely-related work.

\vspace{0.1in}
\noindent {\bf \emph{Machine learning approaches.}}
Most prior malware classification techniques are based on machine learning~\cite{drebin,holmes,smit,aisec,DroidSift,peng12,DroidMiner,Bose08}. The key idea underlying all ML-based approaches is to extract a feature vector summarizing the application's behavior and then train a classifier to predict whether an app is malicious or benign.

Many ML-based approaches generate their feature vectors from a graph representation of the program. For 
instance, \textsc{SMIT}~\cite{smit} and Gascon et al.~\cite{aisec} model programs using call-graphs and use clustering algorithms (e.g., KNN) to group similar applications.

Similar to \textsc{SMIT}, both \massvet~\cite{MassVet} and \textsc{DroidSift}~\cite{DroidSift} use a graph abstraction to represent Android apps. Specifically, \massvet~\cite{MassVet} 
computes graph similarity between a given app ${\mathcal A}$ and the database of existing apps  to determine if ${\mathcal A}$ is the repackaged version of an existing app. As we show in our evaluation, \toolname can achieve better precision and fewer false positives compared to \massvet. Similarly, \textsc{DroidSift} abstracts programs  using an \emph{API dependency graph} and employs graph similarity metrics to cluster applications into different malware families. Similar to \toolname, \textsc{DroidSift} performs multi-label rather than binary classification and  employs a semantics-based approach to resist obfuscation. However, unlike \textsc{DroidSift}, \toolname \ can infer signatures from very few samples and does not need a large training set.  
We tried to compare \toolname against \textsc{DroidSift} but we were not able to reproduce their results using \textsc{DroidSift}'s web service~\cite{DroidSift-tool}. While we have contacted the authors,  the issues with \textsc{DroidSift}'s web service have not been resolved by the time of this submission.

Another related tool is \textsc{Holmes}~\cite{holmes}, which detects Windows malware by  constructing the program's \emph{behavior graph}. Such behavior graphs are constructed dynamically
by analyzing data dependencies in program traces. Given a behavior graph, \textsc{Holmes} then uses  \emph{discriminative subgraph mining}  to extract features that can be used to distinguish malicious applications from benign ones.
\textsc{DroidMiner}~\cite{DroidMiner} and Bose et.al~\cite{Bose08} also use a variant of behavior graphs to abstract 
malware as a set of malicious components and use machine learning for classification. In contrast to these these techniques which require a large number of samples (e.g., \textsc{Holmes} uses 492 samples in their evaluation), \toolname requires as few as two samples to automatically generate malware signatures. Second, our constraint-based approach is guaranteed to generate the optimal signature (in terms of maximizing suspicious behaviors). Finally, we believe that the semantic signatures synthesized by \toolname are easier for security analysts to interpret.

Another state-of-the-art malware detector for Android is \drebin~\cite{drebin}, which  combines syntactic and semantic features into a joint vector space. The syntactic features are obtained from the application's manifest file, while the semantic features are obtained through static analysis. Some of the features used by \drebin are similar to \toolname; for instance, \drebin also extracts data flow information as well as suspicious API calls. As we demonstrate experimentally, \toolname can achieve the same or better precision as \drebin using much fewer samples.

\vspace{0.1in}
\noindent {\bf \emph{Signature-based approaches.}} As mentioned earlier, sig\-na\-tu\-re-based approaches look for explicit patterns to identify instances of a malware family~\cite{Kirin,apposcopy,Hancock,SB05,graphmin,malspec,Yedalog}. These signatures can be either syntactic or semantic, or a combination of both. Our approach extends the applicability of signature-based detectors by automatically synthesizing signatures from a handful of malware samples.

Among signature-based detectors, \apposcopy~\cite{apposcopy} uses semantic signatures to identify Android malware. Specifically, it performs a combination of static taint analysis and inter-component control flow analysis to determine whether a signature matches an Android application.
As mentioned earlier, \toolname is integrated as a plug-in to \apposcopy and can generate signatures in \apposcopy's malware specification language.

\textsc{Kirin}~\cite{Kirin}, which is another signature-based tool,  leverages Android permissions to detect  malware. Specifically, \textsc{Kirin} uses permission patterns to perform binary classification of apps as benign vs. malicious. While \textsc{Fact}~\cite{graphmin} obtains a signature by computing the unweighted maximal common subgraph extracted from dynamic analysis, \toolname computes a weighted \emph{maximally suspicious common subgraph} using static analysis.

\vspace{0.1in}
\noindent {\bf \emph{Zero-day malware detection.}}
Zero-day malware detectors~\cite{RiskRanker,DroidRanger,MassVet} can detect malicious applications that belong to new malware families.  For instance, \textsc{RiskRanker}~\cite{RiskRanker} 
ranks Android applications as high-, medium-, or low-risk depending on the presence of suspicious features, such as certain kinds of function calls. As another example,
\textsc{DroidRan\-ger}~\cite{DroidRanger} uncovers zero-day malware by performing heuristic-based filtering to identify  suspicious behaviors. Some ML-based approaches (e.g.,~\cite{drebin},~\cite{Bose08}) can, in principle, also uncover zero-day malware, even though their detection rate is much higher for instances of known malware families. Even though the primary goal of \toolname is not zero-day malware detection, our experiments in Section~\ref{sec:zero} show that \toolname can nonetheless be successfully used to detect instances of unknown malware families.

\vspace{0.1in} 
\noindent {\bf \emph{Information flow analysis for Android.}} Several tools, including \toolname, use information flow as a component of malware signatures or feature vectors. While information flow does not directly predict malware, \toolname \ can benefit from recent  advances in information flow analysis to improve the quality of its signatures.  Some examples of Android information flow analysis tools include \textsc{FlowDroid}~\cite{flowdroid}, \textsc{AppIntent}~\cite{AppIntent}, \textsc{AppAudit}~\cite{AppAudit}, \textsc{TaintDroid}~\cite{taintdroid}, \textsc{DroidScope}~\cite{droidscope}, \textsc{CHEX}~\cite{chex}, \textsc{Epicc}~\cite{epicc}, \textsc{HI-CFG}~\cite{hicfg} and \textsc{AppContext}~\cite{AppContext}. 

\section{Conclusion}

We have presented a new technique for automatically inferring interpretable semantic malware signatures from a small number of malware samples. Our technique significantly improves the usability of signature-based malware detectors by eliminating the human effort required for writing malware signatures. Furthermore, we show that \toolname’s signature inference algorithm enables approximate signature matching, which is useful both for zero-day malware detection and for making our technique more resilient to behavioral obfuscation.

We implemented our technique in a tool called \toolname, which we evaluated both on malicious apps in the Android Genome Malware Project as well as on benign apps from Google Play. Our experiments show that (i) the signatures automatically synthesized by \toolname are better than manually-written signatures in terms of accuracy and false positives, and (ii) the proposed approximate signature matching algorithm allows detecting zero-day and  behaviorally- obfuscated malware with a very low false positive rate. Our tool is publicly available~\cite{astroid-ui} and can be easily used
by security analysts to synthesize malware signatures from very few samples.

\section*{Acknowledgments}

We would like to thank Thomas Dillig, Martin Rinard, Tao Xie, Eric Bodden, Wei Yang and Ashay Rane 
for their insightful comments. We also thank the anonymous
reviewers for their helpful feedback.

This work was supported in part by NSF Award \#1453386, AFRL Awards \#8750-14-2-0270 and \#8750-15-2-0096, and a Google Ph.D. Fellowship.
The views, opinions, and findings contained in this paper are those of
the authors and should not be interpreted as representing the
official views or policies of the Department of Defense or the U.S. Government.

% \fi
%\IEEEpeerreviewmaketitle

{\small
\bibliographystyle{unsrt}
\bibliography{main}
}

\newcommand*{\ext}{} %comment to remove ext
\ifdefined\ext
%\newpage
\appendix
\section*{Appendix A: Interpretability of Explanations}
\label{sec:interpret}

\begin{figure*}
\centering
\scriptsize
\scalebox{0.98}{
\begin{tabular}{ccc}
$\!\begin{aligned}
V&=\{r:\text{receiver},s:\text{service}\} \\
X&=\{(\texttt{SYSTEM},r),(r,s)\} \\
Y&=\left\{
\!\begin{aligned}
&(\texttt{SYSTEM},r,\text{IntentFilter}(\texttt{BOOT\_COMPLETED})), \\
&(\texttt{SYSTEM},r,\text{IntentFilter}(\texttt{SMS\_RECEIVED})), \\
&(\texttt{SYSTEM},r,\text{IntentFilter}(\texttt{PHONE\_STATE})), \\
&(\texttt{SYSTEM},r,\text{IntentFilter}(\texttt{NEW\_OUTGOING\_CALL})), \\
&(s,s,\text{SuspiciousAPI}(\texttt{sendTextMessage})), \\
&(s,s,\text{TaintFlow}(\text{DeviceID},\text{Internet})), \\
&(s,s,\text{TaintFlow}(\text{SubscriberID},\text{Internet})), \\
&(s,s,\text{TaintFlow}(\text{File},\text{Internet})), \\
&(s,s,\text{TaintFlow}(\text{SimSerial},\text{Internet}))
\end{aligned}
\right\}
\end{aligned}$
&
$\!\begin{aligned}
&\text{SuspiciousAPI}(\texttt{sendSMS})~(1.07) \\
&\text{NetworkAddress}(\texttt{lebar.gicp.net})~(0.93) \\
&\text{Permission}(\texttt{DELETE\_PACKAGES})~(0.58) \\
&\text{IntentFilter}(\texttt{SMS\_RECEIVED})~(0.56) \\
&\text{SuspiciousAPI}(\texttt{getSubscriberID})~(0.53)
&\end{aligned}$
&
$\!\begin{aligned}
&\texttt{onReceive}~\to\texttt{createFromPdu} \\
&\texttt{onReceive}~\to\texttt{getOriginatingAddress} \\
&\texttt{onReceive}~\to\texttt{getDisplayMessageBody}
\end{aligned}$
\\
\hline
\toolname & \drebin & \massvet \\
& &
\end{tabular}
}
\caption{Explanations produced by each tool for the GoldDream family.}
\label{fig:golddreamsig}
\end{figure*}

In this section, we describe the explanations for the GoldDream malware family generated by \toolname, \drebin and \massvet, which are shown in Figure~\ref{fig:golddreamsig}.
In particular, we discuss how a security analyst might use these explanations to pinpoint and understand the malicious behaviors of GoldDream malware.

The left-hand side of Figure~\ref{fig:golddreamsig} shows the signature synthesized by \toolname for the GoldDream malware family. This signature conveys a wealth of information about the malware family to the auditor:

\vspace{0.03in}
\noindent {\bf \emph{Components.}} Simply by looking at the vertices $V$ in the signature, the auditor sees that the malware consists of two components: a receiver $r$ and a service $s$. Furthermore, the inter-component call relations $X$ convey that $r$ is called by the Android framework (denoted by the vertex \texttt{SYSTEM}), and $s$ is subsequently called by $r$.

\vspace{0.03in}
\noindent{\bf \emph{Triggers.}} The metadata $Y$ encodes intent filters registered by each receiver. For the GoldDream family, the receiver $r$ can be triggered by a variety of common system events including \texttt{BOOT\_COMPLETED} (triggered when the app starts) and \texttt{SMS\_RECEIVED} (triggered when an SMS message is received).

\vspace{0.03in}
\noindent{\bf \emph{Malice.}} The metadata $Y$ also encodes malicious behaviors associated with the relevant components. For the GoldDream family, this includes suspicious API calls (e.g., the service $s$ calls \texttt{sendTextMessage}) and information leaks (e.g., the service $s$ leaks the device ID to the Internet).

\vspace{0.03in}
In summary, the signature inferred by \toolname encodes that members of the GoldDream family contain a receiver triggered by common system events, and this receiver calls a service that leaks sensitive information to the Internet.

Unlike \toolname, \drebin and \massvet do not characterize the malice corresponding to a particular malware family. Instead, they produce explanations for why a specific app might be malicious.

\vspace{0.03in}
\noindent{\bf \emph{Comparison to Drebin.}}
The explanation produced by \drebin consists of a list of the features most indicative of malicious behavior together with weights indicating their relative significance. Figure~\ref{fig:golddreamsig} (middle) shows this list of top features and corresponding weights for the GoldDream family (obtained by averaging over all members of the family). The most significant feature is the call to the suspicious API \texttt{sendSMS}. Only 6.6\% of benign apps in the Drebin dataset call \texttt{sendSMS}, but 92.8\% of GoldDream malware make this call. Therefore, calling \texttt{sendSMS} is a good statistical signal that an app is malicious, but it does not give conclusive evidence of malice. Unlike \toolname, \drebin fails to pinpoint any malicious components, intent filters, and information leaks, let alone the complex relationships between these entities.

The presence of the network address \texttt{lebar.gicp.net} might be a more conclusive signal of malice, but if this feature were used to filter apps on Google play, then malware developers would quickly learn to obfuscate it. In contrast, \toolname solely relies on semantic features of apps that are significantly harder to obfuscate.

\vspace{0.03in}
\noindent{\bf \emph{Comparison to MassVet.}}
The explanation produced by \massvet consists of a set of method calls added as part of the repackaging process. These method calls typically do not pinpoint the malicious functionality in the app. For example, the explanation produced by \massvet for an instance of the GoldDream family is shown in Figure~\ref{fig:golddreamsig} (right). It consists of a list of calls to APIs that \massvet considers suspicious. While these API calls play a role in the malicious functionality of this app, they are also commonly used by benign apps to process SMS messages, and do not capture the overall malice present in this app.

\vspace{0.03in}
In addition, we believe the results produced by \toolname confer a number of other benefits:

\vspace{0.03in}
\noindent {\bf \emph{Malware family.}} Unlike \drebin and \massvet, which can only identify whether an app is malicious or benign, \toolname determines which malware family the app belongs to. Since some malware families are more malicious than others, the ability to categorize different apps into malware families provides finer-grained information about the threat level compared binary classification as malicious vs. benign. For instance, some malware
families merely affect user experience~\cite{ikee} whereas others introduce financial risks by stealing user's personal account or credit card information~\cite{fakeneflix}.

\vspace{0.03in}
\noindent {\bf \emph{Disinfection.}} Since \toolname pinpoints the malicious components in the malicious app, it can be used to ``disinfect'' the app by removing these malicious components.

\fi

\end{document}